\documentclass[11pt]{article}

\usepackage{setspace}
\usepackage[T1]{fontenc}
\usepackage[utf8]{inputenc}
\usepackage{lmodern}
\usepackage{textcomp}
\usepackage{fancyhdr}
\usepackage[english]{babel}
\usepackage{graphicx}
\usepackage{tabularx}
\usepackage{amsmath}
\usepackage{amssymb}
 \numberwithin{equation}{section}
\usepackage{amsmath,amssymb,amsfonts}
\usepackage{graphicx}
\usepackage{hyperref}
\usepackage{srcltx}

\begin{document}

\newcommand{\ack}[1]{[{\bf Pfft!: {#1}}]}
\newcommand\bC{\mathbb{C}}
\newcommand\bP{\mathbb{P}}
\newcommand\bR{\mathbb{R}}
\newcommand\bZ{\mathbb{Z}}

\newcommand{\Ccal}{{\cal C}}
\newcommand{\Ocal}{{\cal O}}

\newcommand{\Xtil}{\tilde{X}}

\newcommand\dlangle{\langle\langle}
\newcommand\drangle{\rangle\rangle}
\newcommand{\bra}{\langle}
\newcommand{\ket}{\rangle}
\newcommand{\Rmath}{\mathbb{R}}
\newcommand{\Zmath}{\mathbb{Z}}

\newcommand{\be}{\begin{equation}}
\newcommand{\ee}{\end{equation}}
\newcommand{\bea}{\begin{eqnarray}}
\newcommand{\eea}{\end{eqnarray}}

\newcommand{\bel}[1]{\begin{equation}\label{eq:#1}}
\newcommand{\eq}[1]{(\ref{eq:#1})}
\newcommand{\eref}[1]{(\ref{#1})}

\newcommand{\pa}{\partial}
\newcommand{\mlambda}{{\underline{\underline{\lambda}}}}
\newcommand{\Tr}{\mathrm{Tr}}

\def\str{{str}}
\def\lstr{\ell_\str}
\def\gstr{g_\str}
\def\Mstr{M_\str}
\def\eps{\epsilon}
\def\varep{\varepsilon}
\def\del{\nabla}
\def\grad{\nabla}
\def\tr{\hbox{tr}}
\def\perp{\bot}
\def\half{\frac{1}{2}}
\def\p{\partial}

\renewcommand{\thepage}{\arabic{page}}
\setcounter{page}{1}

\rightline{HIP-2015-31/TH}
\rightline{INT-PUB-15-048}

\vskip 0.75 cm
\renewcommand{\thefootnote}{\fnsymbol{footnote}}
\centerline{\Large \bf Conformal quantum mechanics and holographic quench}
\vskip 0.75 cm

\centerline{{\bf Jarkko J\"arvel\"a${}^{1,2}$\footnote{jarkko.jarvela@helsinki.fi}, Ville Ker\"anen$^{3}$\footnote{vkeranen1@gmail.com} and
Esko Keski-Vakkuri${}^{1,2}$\footnote{esko.keski-vakkuri@helsinki.fi},
}}

\vskip .5cm
\centerline{${}^1$\it Department of Physics,}
\centerline{\it P.O. Box 64, FIN-00014 University of Helsinki, Finland}
\vskip .5cm
\centerline{${}^2$\it Helsinki Institute of Physics,}
\centerline{\it P.O. Box 64, FIN-00014 University of Helsinki, Finland}
\vskip .5cm
\centerline{${}^3$\it Rudolf Peierls Centre for Theoretical Physics, University of Oxford,}
\centerline{\it 1 Keble Road, Oxford OX1 3NP, United Kingdom}
\vskip .5cm

\setcounter{footnote}{0}
\renewcommand{\thefootnote}{\arabic{footnote}}

\begin{abstract}
Recently, there has been much interest in holographic computations of two-point non-equilibrium Green functions from AdS-Vaidya
backgrounds. In the strongly coupled quantum field theory on the boundary, 
the dual interpretation of the background is an equilibration process called a holographic quench. 
The two dimensional AdS-Vaidya spacetime is a special case, dual to conformal quantum mechanics. We study how the quench is
incorporated into a Hamiltonian $H+\theta(t) \Delta H$ and into correlation functions.
With the help of recent work on correlation functions in conformal quantum mechanics, we
first rederive the known two point functions, and then compute non-equilibrium 3- and 4-point functions.
We also compute the 3-point function Witten diagram in the two-dimensional AdS-Vaidya background,
and find agreement with the conformal quantum mechanics result.
\end{abstract}

\newpage

\section{Introduction and summary}

Partially motivated by the AdS/CFT correspondence, there has been recent  progress in understanding correlation functions in conformally invariant quantum
mechanics (CQM), relevant for the AdS$_2$/CFT$_1$ case. In this case, the $SO(2,1)$-isometry of the bulk AdS$_2$ manifold manifests as
a $SO(2,1)$ conformal invariance of a quantum mechanical theory on the boundary.

The benchmark model of conformal quantum mechanics, viewed as the $D=1$ limit of $SO(D+1,1)$ conformally
invariant scalar field theories with a $g\phi^{2D/(D-2)}$ potential in $D$ spacetime dimensions, was studied by de Alfaro, Fubini and Furlan (dAFF) \cite{daff}. 
The action is invariant under $SO(2,1)$ {\em i.e.} $SL(2,\mathbb{R})$ transformations,
with the generators satisfying, after quantization, the commutation relations
\begin{equation}\label{sl2r}
i[D,H]=H,\quad i[D,K]=-K,\quad i[K,H]=2D
\end{equation}
of a $so(2,1)$ Lie algebra. There are many other quantum mechanical realizations of (\ref{sl2r}) besides the original dAFF model.
For the AdS$_2$/CFT$_1$ correspondence, we expect the relevant model to have many ($N$) interacting degrees of freedom, with the bulk calculations
corresponding to a large $N$, strong coupling limit. However, the full picture of the AdS$_2$/CFT$_1$ correspondence is not yet well understood, due to complications
arising from the AdS$_2$  fragmentation and backreaction when the geometry is reached from string theory in a form 
AdS$_2\times X$ where $X$ is a compact space \cite{Maldacena:1998uz}.  Recently, these issues were studied in context of 1+1 dimensional models  of dilaton gravity coupled to 
matter \cite{Almheiri:2014cka}. There one can also construct solutions for an ingoing null matter pulse into a vacuum, creating a black hole, so that the two-dimensional
spacetime is described by the AdS-Vaidya metric. 

The null collapse to a black hole, or the AdS-Vaidya spacetime has been much studied as an analytic holographic model of one type of a global quench in the QFT on the boundary (see \cite{Banks:1998dd} for
an early idea, and {\em e.g.} \cite{Hubeny:2007xt,AbajoArrastia:2010yt,Albash:2010mv,Balasubramanian:2010ce,Balasubramanian:2011ur} for early papers on
AdS-Vaidya and holographic quench). In particular, the model
enables one to study nonequilibrium Green functions for a strongly coupled QFT, as these can be computed from the bulk geometry. However, it is not very transparent how exactly
the quench is realized on the boundary. Furthermore, especially in higher dimensions the Green functions do not usually have a simple analytic form. In this paper, we are interested
in studying the holographic quench in 0+1 dimensions, and the structure of the Green functions. We work in a simplified setting, assuming that the bulk spacetime is described by
the two-dimensional AdS-Vaidya metric. For the dual theory at the boundary, we rely on the recent work \cite{cjps}, where 
the authors studied how to construct correlation functions in conformal quantum mechanics.
Using only the $so(2,1)$ algebra, without specifying the underlying theory, they constructed operators which satisfied some,
but not all properties of conformal primaries, and a vacuum state which was not fully invariant under (\ref{sl2r}). However, despite these shortcomings
when the vacuum and the operators are combined to correlation functions, the result satisfies expected
transformation properties for a conformally invariant theory.

In addition to the general motivation of extending the study of quenches modeled by AdS-Vaidya spacetimes to the AdS$_2$/CFT$_1$ case, we have specific motivations 
and goals: 1) it is interesting to study to what extent non-equilibrium correlation functions can be computed working directly on the boundary theory; 2) we would like
to understand in more detail what kind of a quench the AdS-Vaidya bulk spacetime corresponds to in the boundary; 3) 3-pt and 4-pt correlation functions have been studied at thermal equilibrium 
\cite{3pt4pt} and more recently in the context of semiclassical limits by bulk geodesic methods \cite{Fitzpatrick:2014vua,Asplund:2014coa,Hijano:2015rla}, but not yet in the context of global quenches.  
In this work we interpret
the AdS$_2$-Vaidya spacetime to be realized in CQM as a sudden change of the Hamiltonian. We compute  
analytic results for thermal and non-equilibrium 2-pt, 3-pt and 4-pt functions in quenched conformal quantum mechanics, by using the results of  \cite{cjps}, and the 2-pt and 3-pt functions also from a bulk AdS$_2$-Vaidya background calculation, finding agreement with the results from the CQM computation.  

This paper is organized as follows: Section 2 reviews some key concepts of conformal quantum mechanics and various choices of time evolution; in Section 3 we introduce
the AdS$_2$-Vaidya spacetime as a model of a quench, which we find to be realized in CQM as a sudden change of the Hamiltonian, and 
then compute non-equilibrium two-point and 3-point correlation
functions both by the holographic method from the bulk spacetime and by a direct calculation in CQM. Section 4 extends the CQM calculation to the 4-point function, and ends with
some brief comments.

\section{Review}

We first review some key features of $so(2,1)$ representation theory.
The energy eigenvalues (of $H$) are continuous and the eigenstates are all non-normalizable.
In addition, there is no state that would vanish under all three generators i.e. that would be invariant under all the symmetry transformations.
Let us then consider another complete set of orthonormal states. Moving to a new basis of operators with the commutators
\begin{eqnarray}
R &=& \frac{1}{2}\left(\frac{K}{a}+aH\right), \nonumber \\
L_{\pm} &=& \frac{1}{2}\left(\frac{K}{a}-aH\right)\pm i D \ ,
\end{eqnarray}
where $a$ is a constant with the dimension of time\footnote{The constant $a$ played the role of an infrared regulator in \cite{daff}.}, the commutation relations become
\be
[ R,L_{\pm}] = \pm L_{\pm},\quad [L_-,L_+]=2R \ ,
\ee
so that they form the Cartan-Weyl basis of the $so(2,1)$ algebra.
 The generator $R$ is compact with a discrete set of eigenstates.
The discrete lowest weight representation can then be constructed using $L_{\pm}$ as raising and lowering operators. Explicitly:
\begin{eqnarray}
R|n\rangle &=& r_n|n\rangle,\\
r_n&=&r_0+n
\ ,\,\,\, n\in\mathbb{N}\\
\langle n'|n\rangle &=&\delta_{n',n}, \\
L_{\pm}|n\rangle &= &\sqrt{r_n(r_n\pm 1)-r_0(r_0-1)}|n\pm 1\rangle, \mbox{ and}\\
|n\rangle &= &\sqrt{\frac{\Gamma(2r_0)}{n!\Gamma(2r_0+n)}}(L_+)^n|0\rangle \ ,
\end{eqnarray}
where $r_0$ is a parameter, the lowest weight of the representation, also connected to the eigenvalue $r_0(r_0-1)$ of the
Casimir invariant $\mathcal{C}=\frac{1}{2}(HK+KH)-D^2=R^2-L_{+}L_-$  of the algebra\footnote{The value of $r_0$ depends on the theory. For
example, in the inverted harmonic oscillator model introduced in \cite{daff}, $r_0$ is related to the value of the 
dimensionless coupling constant $\lambda$ in the potential term $-\lambda q^{-2}$. }. In particular, the lowest eigenstate $|0\ket \equiv |n=0\ket$ plays
the role of a vacuum, called the $R$-vacuum.
For a generic $r_0$, the eigenvalue equation implies that the $R$-vacuum cannot be annihilated by all the $so(2,1)$ generators.

One can also construct a continuous basis of states, $|\tau \rangle$. Using the conventions in \cite{daff,cjps}, the operators take the form
\begin{eqnarray}
H &=& -i\frac{d}{d\tau}\\
D &=& -i(\tau\frac{d}{d\tau}+r_0)\\
K &=& -i(\tau^2\frac{d}{d\tau}+2r_0\tau).
\end{eqnarray}
Constructing the operator $R$ in this basis, one can relate the $|n\rangle$ and $|\tau \rangle$ bases with the help of a differential equation:
\begin{eqnarray}
&\langle \tau|R|n\rangle = r_n\langle \tau |n\rangle = \frac{i}{2}\left[\left(a+\frac{\tau^2}{a}\right)\frac{d}{d\tau}+2r_0\frac{\tau}{a} \right]\langle \tau |n\rangle& \nonumber\\
&\Rightarrow \beta_n(\tau) \equiv \langle \tau |n\rangle = (-1)^n\left[\frac{\Gamma(2r_0+n)}{n!}\right]^{\frac{1}{2}}\left(\frac{a-i\tau}{a+i\tau}\right)^{r_n}\frac{1}
{\left(1+\frac{ \tau^2}{a^2}\right)^{r_0}}&
\end{eqnarray}
The constants in the expression are determined by the condition that the raising and lowering operators operate in the same way in both bases.

The $|\tau\rangle$ basis is not orthonormal, but the overlap of two states has the form
\begin{equation}\label{gf}
\langle \tau_1 |\tau_2\rangle = \sum\limits_{n=0}^{\infty}\beta_n(\tau_1)\beta_n^*(\tau_2) = \Gamma(2r_0)\left(\frac{a}{i2(\tau_1-\tau_2)}\right)^{2r_0} \ ,
\end{equation}
the same as for primary operators of conformal weight $r_0$ in a conformal theory.
In \cite{cjps},
the authors gave an explicit interpretation of (\ref{gf}) as a vacuum 2-point function, constructing an operator $O(\tau)$ which reproduces the $|\tau\rangle$ states
operating on the R-vacuum, $|\tau\rangle = O(\tau)|0\rangle$. Explicitly,
\be
O(\tau) = N(\tau)\exp(-\omega(\tau)L_+) \ ,
\ee
with
\begin{eqnarray}
N(\tau) &=& [\Gamma(2r_0)]^{1/2}\left(\frac{\omega(\tau)+1}{2}\right)^{2r_0} \ , \nonumber \\
\omega(\tau) &=& \frac{a+i\tau}{a-i\tau} \ .
\end{eqnarray}
so that
\be
\langle \tau_2 |\tau_1\rangle =\langle 0|O^{\dagger}(\tau_2)O(\tau_1)|0\rangle \ .
\ee

There are some issues with this form. While it does produce the correct state $|\tau\rangle$, the action of $O^{\dagger}$ on the $R$-vacuum just returns the R-vacuum
multiplied with the normalization factor $N(\tau)$. This would cause the time-translation invariance to be lost when considering $\langle 0|O(\tau_1)O^{\dagger}(\tau_2)|0\rangle$.
One could consider an alternative definition for $O$ which would still produce the same $|\tau\rangle$ states\footnote{There are some other remaining issues too. It can be shown that $\exp[iH\Delta \tau]|\tau\rangle=|\tau+\Delta \tau\rangle$, but $\exp[iH\Delta \tau]O(\tau)\exp[-iH\Delta \tau]\neq O(\tau+\Delta \tau)$ which
makes the Heisenberg picture a bit problematic. In addition, the R-vacuum expectation value  $\bra 0|O(\tau )|0 \ket$ does not vanish, unlike what happens in CFT
for a primary operator -- the R-vacuum is not conformally invariant.}. We could use
\begin{eqnarray*}
O(\tau) = N(\tau)\exp[-\omega(\tau)L_+]\exp[-\omega^*(\tau)L_-]
\end{eqnarray*}
With this definition, $O(\tau)$ would produce the same $|\tau\rangle$ states but this time $O^{\dagger}(\tau)$ produces the state $|\tau\rangle^*$, i.e. a state with complex conjugated coefficients. This operator is almost Hermitian, apart from the complexity of $N(\tau)$. This alternative definition is not necessary for the calculations in this paper.

However, there is another alternative form, suggested in \cite{Jackiw:2012ur}. They suggested
\be
O(\tau) = \frac{[\Gamma(2r_0)]^{1/2}}{2^{2r_0}} e^{i\tau H } e^{aH}
\ee
as it correctly reproduces the $|\tau\rangle$ state when acting on $|n=0\rangle$. The leftmost exponential gives the time evolution from the $|\tau =0\ket$ state which the 
rightmost exponential prepares from the R-vacuum.

\subsection{Alternative time evolutions}\label{sec:timeevolve}

So far we have chosen $H = -i \frac{d}{d\tau}$ as the time evolution generator. Alternative choices have been considered previously in
\cite{Jackiw:2012ur,hartnoll,nakayama} and this enables one to compute Green functions periodic in imaginary time,  associated with a finite temperature background.
We motivate the alternative time evolution generators by reviewing the isometry of two-dimensional anti-de Sitter spacetime AdS$_2$.

Three metrics are often used in the context of AdS$_2$: the Poincar\'e metric (the metric written in the Poincar\'e coordinate patch covering a
part of the spacetime manifold), the global metric (using the global coordinates patch that covers the full AdS covering space manifold), and the black
hole metric (using a coordinate patch, which after periodic identification in imaginary time defines an anti-de Sitter black hole) (see \cite{spradlin-strominger} and Figure 3 therein
for an illustration of the patches). The metrics have the form
\be
{\rm Global:} \ \ \  ds^2 = \frac{R^2}{\cos^2 \theta} (-dT^2 + d\theta^2 ),\,\, -\infty<T<\infty,\,\, -\frac{\pi}{2}<\theta<\frac{\pi}{2}
\ee
\be
 {\rm Poincar\'e:} \ \ \ ds^2 = \frac{R^2}{z^2} (-d\tau^2 + dz^2),\,\, -\infty<\tau<\infty,\,\, 0<z<\infty
\ee
\be\label{bhmetric}
{\rm Black\ hole:} \ \ \ ds^2 = \frac{R^2}{u^2}\left( -\left(1-\frac{u^2}{u_H^2}\right)dt^2 +\frac{du^2}{\left(1-\frac{u^2}{u_H^2}\right)}\right),\,\, -\infty<t<\infty,\,\, 0<u<u_H
\ee
and the coordinates are related by the transformations
\bea
 \tau\pm z &=& 2a\,\tan \left[(T\pm \theta \pm \frac{\pi}{2})/2\right],\\
z&=&\frac{2 u}{1+\cosh(v/u_H)+(u/u_H)\sinh(v/u_H)},
\quad
\tau-z=2 u_H \tanh\left(\frac{v}{2u_H}\right),\\
v(u,t) &=& t-u_H\,{\rm artanh}\left(\frac{u}{u_H}\right).
\eea
Here, we used different symbols for the variables of different choices of metrics for the sake of clarity.

The Killing vector fields that generate the $SO(2,1)$ isometry of AdS$_2$ are represented in Poincar\'e coordinates by
\begin{eqnarray}
 && H = -i \partial_\tau \nonumber \\
 && D = -i \left( \tau\partial_\tau + z\partial_z \right) \nonumber \\
 && K = -i \left[\left( \tau^2 + z^2 \right)\partial_\tau + 2\tau z \partial_z\right]
\end{eqnarray}

After transforming to global coordinates, the Killing vector that generates time translations in global time $T$
is the linear combination (see  also \cite{Ho:2004qp})
\be\label{HG}
H_G =  aH + \frac{K}{4a}  = -i\partial_T \ .
\ee
Likewise, after transforming to black hole coordinates, time translations in the time $t$ are generated by the combination
\be\label{HBH}
H_{BH} =  H - K/(2 u_H)^2  = -i\partial_t \ .
\ee
On the (conformal) boundary of the spacetime, the relation between the different time coordinates reduces to
\be
 \tau = 2a\,\tan (T/2) = 2 u_H\tanh (t / 2u_H)
\ee
The linear combinations (\ref{HG}) and (\ref{HBH}) are the alternative time-evolution generators considered in \cite{Jackiw:2012ur}. Moving to conformal quantum mechanics, the generators are represented as infinitesimal translation operators when acting on the time basis states, {\it e.g.} $H|\tau \ket = -i\frac{d}{d\tau}|\tau \ket$. For the generators $H_G,H_{BH}$, one needs to define new time states $|T\ket, |t\ket$ \cite{Jackiw:2012ur}.
Since $|\tau \ket = O(\tau)|0\ket$, and (in correlation functions) $O(\tau )$ transforms under coordinate transformation like a primary of weight $r_0$, one
defines new time states $|\tilde{t}\ket$ by
\be
  | \tilde{t} \ket = \left( \frac{dt}{d\tilde{t}}\right)^{r_0} |t=t(\tilde{t})\ket \ ,
\ee
or in our case explicitly:
\bea
  && |T\ket = a^{r_0}(\cos(T/2))^{-2r_0} |\tau= 2a\tan (T/2)\ket \\
  && |t \ket = (\cosh (t /(2 u_H)))^{-2r_0} |\tau=2 u_H\tanh (t /(2u_H))\ket \ .
\eea
The different choices of a Hamiltonian
are then represented in the different time state basis as
\bea
   && H_P|\tau\ket = -i \frac{d}{d\tau} |\tau \ket \\
    && H_G|T \ket = -i \frac{d}{dT} |T\ket \\
     && H_{BH}|t\ket = -i \frac{d}{dt} |t\ket \ ,
\eea
inherited from the time evolutions in the three customary AdS$_2$ coordinate patches. 
The two-point function
$ G_2(\tau_2,\tau_1) = \bra \tau_2 | \tau_1 \ket $
can then be related to two-point functions $\bra T_2| T_1\ket$ and $\bra t_2|t_1 \ket$ in terms of a simple rule. Although
the operator $O(\tau )$ is not a primary one, it transforms as one in correlation functions evaluated with respect to the $R$-vacuum
state $|0\ket \equiv |n=0\ket$,
\be\label{conftrans}
 \bra \tilde{t}_2 | \tilde{t}_1\ket \equiv \bra 0 | \tilde{O}^\dagger (\tilde{t}_2) \tilde{O}(\tilde{t}_1) | 0 \ket
 = \bra 0 |  \left( \frac{dt_2}{d\tilde{t}_2}\right)^{r_0} O^\dagger (t_2(\tilde{t_2}))
 \left( \frac{dt_1}{d\tilde{t}_1}\right)^{r_0} O (t_1(\tilde{t_1})) |0 \ket \ .
\ee
This leads to
\bea
\bra T_2| T_1\ket = \bra 0 | O^\dagger_G (T_2) O_G(T_1) | 0 \ket = \frac{\Gamma (2r_0)a^{2r_0}}{[4 i\sin (\frac{T_2-T_1}{2})]^{2r_0}}
\nonumber \\
\bra t_2| t_1\ket = \bra 0 | O^\dagger_{BH} (t_2) O_{BH}(t_1) | 0 \ket = \frac{\Gamma (2r_0)a^{2r_0}}{[4 u_Hi\sinh (\frac{t_2-t_1}{2 u_H})]^{2r_0}} \  .
\eea

\section{AdS-Vaidya geometry and holographic quench}

We now move to consider a holographic model of a quench, and incorporate it into conformal quantum mechanics. In gauge-gravity duality, a quench in the strongly coupled
theory on the boundary has a holographic dual interpretation in the bulk AdS geometry. Perhaps the simplest and most popular model that has been studied is
the AdS-Vaidya spacetime. It describes lightlike collapse of matter into a black hole in AdS$_{D+1}$ space. The holographic dual
interpretation is that the theory on the $D$ spacetime dimensional boundary is initially in a vacuum state, then is at $t=0$ instantaneously sourced in a
homogeneous manner into an excited state which then time evolves into thermal equilibrium. The initial non-equilibrium state
is somewhat special, because expectation values of all local operators thermalize instantaneously. The non-equilibrium nature is only
revealed by considering expectation values of non-local operators, such as 2-point functions or Wilson loops. The two-dimensional AdS-Vaidya spacetime is even more
special, because the boundary has no space directions. In this paper, we are intereted in correlation functions. 2-point (autocorrelation) functions in AdS$_2$-Vaidya background have
been computed in \cite{Ebrahim:2010ra,Keranen:2014lna}. We rederive the result by a simpler calculation using a coordinate transformation. We will then show that it is very simple to obtain in conformal
quantum mechanics.
\begin{figure}[t] 
\begin{center}
\vspace{-0.cm}
\hspace{-.5cm}
\includegraphics[width=.46\textwidth]{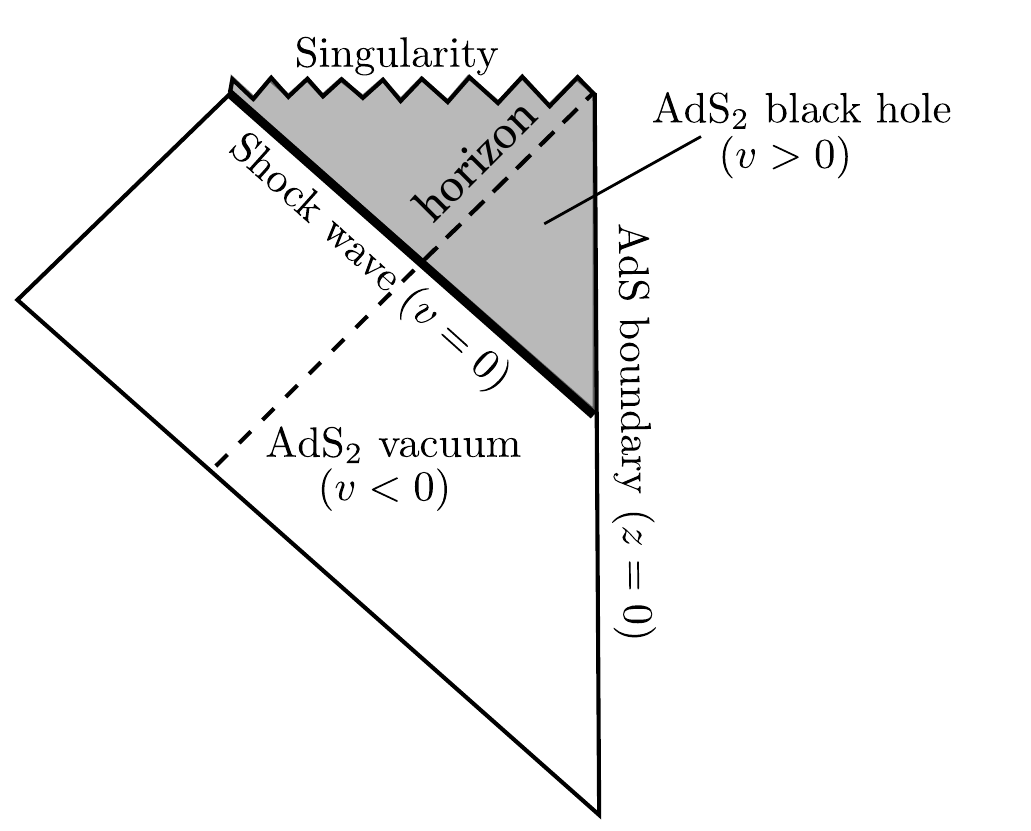}
\end{center}
\vspace{-.3cm}
\caption{\label{fig:adsvaidya}
Penrose diagram of the AdS$_2$-Vaidya spacetime. 
}
\end{figure}

The AdS$_2$-Vaidya spacetime (the Penrose diagram is depicted in Figure \ref{fig:adsvaidya}, see also \cite{Almheiri:2014cka}, Figure 1,  for a related spacetime illustration)  is specified by the metric
\be
ds^2=\frac{1}{z^2}\Big[-\Big(1-\theta(v)\frac{z^2}{z_H^2}\Big)dv^2-2dv dz\Big].\label{eq:vaidyametric}
\ee
For $v<0$, the metric (\ref{eq:vaidyametric}) is the metric of AdS$_2$ written in terms of a lightcone time coordinate $v=\tau-z$.
For $v>0$, the metric  (\ref{eq:vaidyametric}) is the metric of an AdS$_2$ black hole spacetime, with $dv=dt-dz/(1-(z/z_H)^2)$ (we used $u$ instead of $z$ in (\ref{bhmetric})). It is useful to note that the black hole spacetime metric can
also be transformed to the AdS$_2$ metric with a change of coordinates (that also acts on the boundary)
\be
\bar{z}=\frac{2 z}{1+\cosh(v/z_H)+(z/z_H)\sinh(v/z_H)},
\quad
\bar{v}=2 z_H \tanh(v/(2z_H)).
\ee

We can also combine the vacuum and black hole regions of the Vaidya spacetime by a continuous piecewise coordinate transformation
\be
\bar{z}= \left\{\begin{array}{cl}
z, & v<0\, \\
\frac{2 z}{1+\cosh(v/z_H)+(z/z_H)\sinh(v/z_H)}, & v>0,\\
\end{array} \right.\label{eq:coordinate1}
\ee
and
\begin{equation}
\bar{v}= \left\{\begin{array}{cl}
v, & v<0\, \\
2 z_H \tanh(v/(2z_H)), & v>0,\\
\end{array} \right.\label{eq:coordinate2}
\end{equation}
so that the AdS$_2$-Vaidya metric reduces globally to that of AdS$_2$,
\be
ds^2=\frac{1}{\bar{z}^2}\Big(-d\bar{v}^2-2d\bar{v}d\bar{z}\Big).
\ee
For correlation function calculations in the AdS-Vaidya background, it is simple to work with the $(\bar{z},\bar{v})$ coordinates.
Final results are then obtained by performing the inverse coordinate transformation in the end.

\subsection{AdS-Vaidya 2-point function}

The (near boundary) vacuum two point function in AdS$_2$ is well known. Let us consider for simplicity a massless bulk scalar field, 
\be
G_F^{(2)}(\bar{z}_1,\bar{v_1};\bar{z}_2,\bar{v}_2)\approx\frac{1}{\pi}\frac{\bar{z}_1 \bar{z}_2}{-(\bar{v}_2-\bar{v}_1)^2+i\epsilon} \ .\label{Vaidyaboundary}
\ee
We then perform the inverse coordinate transformation to the original coordinates $v,z$ and obtain
the in-in (vacuum) two point function in the AdS$_2$-Vaidya. Note that
the bulk Klein-Gordon equation of motion uniquely fixes the
full two point function with the initial condition that it must for $\bar{v}_1<0$ and $\bar{v}_2<0$ agree with the vacuum
two point function. Clearly this is the case for (\ref{Vaidyaboundary}).

Let us then consider the region $\bar{v}_2 >0$ and $\bar{v}_1<0$. Then, we  use the inverse coordinate transformation, and take the boundary limit with the scaling
prefactors \cite{Banks:1998dd} to find the boundary two point function for an operator with scale dimension $\Delta =1$.
\begin{align}
G^{(2)}_{F,\partial}(t_2,t_1)&=\lim_{z_1,z_2\rightarrow 0}(z_1 z_2)^{-1}G_F(z_1,t_1;z_2/\cosh^2\frac{t_2}{2z_H},2 z_H\tanh\frac{t_2}{2 z_H})\nonumber
\\
&=\frac{1}{\cosh^2\frac{t_2}{l}}\frac{1}{\pi}\frac{1}{-(t_1-l\tanh\frac{t_2}{l})^2+i\epsilon}\nonumber
\\
&=\frac{1}{\pi}\frac{1}{-(t_1\cosh\frac{t_2}{l}-l\sinh\frac{t_2}{l})^2+i\epsilon},
\end{align}
where we use the notation $l=2z_H$. From now on, we will be using the parameter $l$ which gives us the temperature $T=1/(\pi l)$. This result was obtained earlier in \cite{Ebrahim:2010ra,Keranen:2014lna,spectral} using more complicated methods.
For $v_2>0$ and $v_1>0$, we simply
obtain the thermal two point function
\be
G^{(2)}_{F,\partial}(t_2,t_1)=\frac{1}{\pi}\frac{1}{-(l\sinh\frac{t_1-t_2}{l})^2+i\epsilon}.
\ee
Similarly, for an operator with a generic scale dimension $\Delta \neq 1$, the exponent in the denominator changes, {\em e.g.} (\ref{Vaidyaboundary}) becomes
\be\label{2ptfrombulk}
G^{(2)}_{F,\partial}(t_2,t_1)=\frac{1}{\pi}\frac{1}{\left[-(t_1\cosh\frac{t_2}{l}-l\sinh\frac{t_2}{l})^2+i\epsilon\right]^\Delta} \ .
\ee

\subsection{Holographic quench and correlation functions in conformal quantum mechanics}

The AdS-Vaidya spacetime corresponds to a quench in the boundary theory, but its precise realization has not been transparent. From the discussion in Section (\ref{sec:timeevolve}) we
learn that the two-dimensional AdS-Vaidya bulk geometry corresponds to a sudden change of the time evolution of the system, thus in conformal quantum mechanics the quench corresponds
to the Hamiltonian
\be
           H = H_P + \theta (t) \Delta H \ ,\label{eq:quenchham}
\ee
where $H_P$ generates time evolution with respect to the Poincar\'e time $\tau$ and $\Delta H = -K/(2u_H)^2$, so that after the quench the Hamiltonian becomes $H_{BH}$,
generating time evolution with respect to the black hole time coordinate $t$. The two time coordinates have a common origin $\tau=t=0$ at the quench.

Further, in 1+1 dimensional conformal field theory with a quench, it is non-trivial to compute non-equilibrium correlation functions \cite{Calabrese:2009qy}.
On the other hand, in simple free quantum mechanical
systems such as the harmonic oscillator, one usually works in the Heisenberg picture where in-in vacuum correlation functions can be calculated by working out the appropriate
Bogoliubov transformation.  In conformal quantum mechanics, at least for the holographic quench of interest here, 
the situation is simpler since we do not need the explicit form of the Hamiltonian of the system which would be needed to
evaluate the Bogoliubov transformation. We only need to apply the rule how the operators (or more specifically, correlation functions) transform under a conformal transformation.
For example, consider the non-equilibrium two-point function with $t_2>0$, the operator $O(t_2)$ is inserted after the quench, and $t_1<0$, insertion before the quench. For the insertions,
we define\footnote{Our justification for this definition is the following.  
Recall that the operator $\mathcal{O}(t)$ is defined through the state $|t\rangle=\mathcal{O}(t)|0\rangle$. We
are lead to solve the non-equilibrium quench problem by solving the time dependent Schr\"odinger equation
\be
-i\partial_t|t\rangle= H|t\rangle,\label{eq:quenchSeq}
\ee
where $H$ is the quenched Hamiltonian (\ref{eq:quenchham}). In the Appendix, we show that (\ref{eq:quenchSeq}) is solved by $|t\ket = \mathcal{O}_{Vaidya}(t)|0\rangle$.
}
\be\label{OVaidya}
 O_{Vaidya} (t) = \left\{ \begin{array}{l} O_{BH}(t) \ , \ {\rm for}\ t>0 \\
                                                                  O_{P}(\tau) \ , \ {\rm for}\ t=\tau<0 \end{array} \right. \ , 
\ee
and recall that $\tau = \tanh (t/l)$ for $t>0$, $\tau=t$ for $t<0$.  We then use
the operator $O_{Vaidya}(t)$ for the two insertions. Thus, in order to calculate the correlation functions, we only need to apply the rule (\ref {conftrans}) how the operators (in fact, correlation functions) transform under a conformal transformation, to arrive at
\be
 \bra 0 | O^\dagger_{Vaidya} (t_2) O_{Vaidya}(t_1) | 0 \ket       = \Gamma(2r_0)\left[\frac{a}{2i[t_1\cosh\left(\frac{t_2}{l}\right)-l\sinh\left(\frac{t_2}{l}\right)]}\right]^{2r_0} \ ,
\ee
The result is in agreement (after adjusting the overall normalization) with the above bulk calculation (in the limit $\epsilon=0$), when we match the lowest weight with the scale dimension, $r_0 = \Delta$.

\subsection{Three-point functions}
In conformal quantum mechanics, it is equally straightforward to compute the three- and four-point functions in the holographic quench background. This is a new result, since
previous studies of holographic quenches have been focusing on two-point functions. 
We consider the three-point functions which have already been calculated in the zero temperature conformal quantum mechanics \cite{cjps,Jackiw:2012ur}:
\be
\langle \tau_2|\phi(\tau) |\tau_1\rangle \equiv \langle 0| O(\tau_2)\phi(\tau) O(\tau_1)|0\rangle = \frac{A}{(\tau_1-\tau_2)^{2r_0-\delta}(\tau-\tau_1)^{\delta}(\tau-\tau_2)^{\delta}},
\ee
where $\phi$ is a primary operator with scale dimension $\delta$ and $A$ is a constant,
\be
A= \langle 0|\phi(0)|0\rangle\left(\frac{i}{2}\right)^{2r_0+\delta}\Gamma(2r_0)a^{2r_0}.
\ee
(Note that $ \langle 0|\phi(0)|0\rangle$ needs not vanish, as the R-vacuum is not invariant under all conformal transformations.)

Making a coordinate transformation, $t = l\, \text{artanh}\left(\frac{\tau}{l}\right)$,  and using the conformal transformation rules in the correlator, 
the three-point function becomes
\be
{}\langle t_2|\phi (t) |t_1\rangle = \frac{A}{l^{2r_0+\delta}\sinh^{\delta}\left(\frac{t-t_2}{l}\right)\sinh^{\delta}\left(\frac{t-t_1}{l}\right)\sinh^{2r_0-\delta}\left(\frac{t_1-t_2}{l}\right)},
\ee
which is the expected finite temperature result.

Now, we do a quench at $t=0$, {\em e.g.} we turn on the temperature at $t=0$, and compute the three-point function with $t_2>0$, $t_1<0$. First, we consider $t<0$. In this case,
\bea
{}\langle t_2|\phi(t) |t_1\rangle &=& \frac{A}{(t-t_1)^{\delta}\left(t-l\tanh\left(\frac{t_2}{l}\right)\right)^{\delta}\left(t_1-l\tanh\left(\frac{t_2}{l}\right)\right)^{2r_0-\delta}\cosh^{2r_0}(\frac{t_2}{l})}\\
&=& \frac{A}{(t-t_1)^{\delta}\left(t\cosh(\frac{t_2}{l})-l\sinh\left(\frac{t_2}{l}\right)\right)^{\delta}\left(t_1\cosh(\frac{t_2}{l})-l\sinh\left(\frac{t_2}{l}\right)\right)^{2r_0-\delta}}.
\eea
In the denominator of the last expression, we see the familiar factors from the thermalizing two-point function.

If $t,t_2>0, t_1<0$, the 3-point function becomes
\bea
&& \langle t_2|\phi (t) |t_1\rangle = \mbox{} \\
&&\mbox{} = \frac{A}{(l\tanh\left(\frac{t}{l}\right)-t_1)^{\delta}\left(l\tanh\left(\frac{t}{l}\right)-l\tanh\left(\frac{t_2}{l}\right)\right)^{\delta}\left(t_1-l\tanh\left(\frac{t_2}{l}\right)\right)^{2r_0-\delta}\cosh^{2r_0}(\frac{t_2}{l})\cosh^{2\delta}(\frac{t}{l})}\nonumber \\
&& \mbox{}= \frac{f}{l^{\delta}\left(l\sinh(\frac{t}{l})-t_1\cosh(\frac{t_2}{l})   \right)^{\delta}\sinh^{\delta}\left(\frac{t-t_2}{l}\right) (\cosh(\frac{t_2}{l})t_1-l\sinh(\frac{t_2}{l}))^{2r_0-\delta}}. \nonumber
\eea

We will next compare these results against a holographic derivation of  3-point functions from the AdS$_2$-Vaidya background. We begin by reviewing some facts of the calculation
in an AdS$_2$ vacuum background -- since we are interested in in-in vacuum correlation functions, we need to adopt a Keldysh contour method, which we present next.
(Note: we focus only on the leading contribution to the 3-point function, ignoring the additional contributions associated with backreaction which could be computed if bulk
1+1 (dilaton) gravitational dynamics would be included as in \cite{Almheiri:2014cka}.)

\subsection{AdS$_2$ vacuum 3-point functions with Keldysh contour}

As in our previous bulk analysis, to keep matters simple we consider a massless scalar field. For the three-point function we add self-interaction terms and  start from the action
\be
S=\int d^2x\sqrt{|g|}\Big(-\frac{1}{2}(\nabla\phi)^2-\frac{\lambda}{3!}\phi^3+...\Big).
\ee
The bulk to bulk Feynman propagator of $\phi$ is given by
\begin{align}
G_F^{(2)}(x_1,x_2)&=-\frac{1}{4\pi}\log\Bigg[\frac{-(\tau_2-\tau_1)^2+(z_1-z_2)^2+i \epsilon}{-(\tau_2-\tau_1)^2+(z_1+z_2)^2+i\epsilon}\Bigg]
\nonumber
\\
&=\frac{1}{\pi}\frac{z_1 z_2}{-(\tau_2-\tau_1)^2+z_1^2+i\epsilon}+\Ocal(z_2^3),
\end{align}
where we denote $x_j=(z_j,\tau_j)$. In the end we are interested in the limit $z_2\rightarrow 0$ and thus, we will work with the order $\Ocal(z_2)$ term above.
Also, we will in the following need the Wightman two-point function, which can be obtained from the Feynman one
using the identity $\langle \phi (x_1)\phi(x_2)\rangle=\theta(\tau_2-\tau_1)\langle T(\phi(x_1)\phi(x_2))\rangle+\theta(\tau_1-\tau_2)\langle T(\phi(x_1)\phi(x_2))\rangle^*$, and is given by
\be
G_+^{(2)}(x_1,x_2)=\frac{1}{\pi}\frac{z_1 z_2}{-(\tau_2-\tau_1)^2+z_1^2+i(\tau_2-\tau_1)\epsilon}+O(z_2^3).
\ee
The 3-point function can be now calculated using perturbation theory in $\lambda$. The calculation would be simplest in
Euclidean time, but since we will later consider a non-equilibrium situation,
which is inherently real time, we will show how to calculate the vacuum 3-point function in the bulk real time formalism.
The boundary correlator is obtained as a limit of the bulk correlator by using the extrapolate dictionary.

The bulk 3-point function, defined in terms of the Heisenberg picture field operators $\phi_H$, can be written in terms of the Dirac/Interaction picture operators $\phi$, using a complex 
time  contour as
\be
G_F^{(3)}(x_1,x_2,x_3)=\langle T(\phi_H(x_1)\phi_H(x_2)\phi_H(x_3))\rangle=\langle T_C\Big( \phi(x_1)\phi(x_2)\phi(x_3) e^{-i\frac{\lambda}{3!}\int_C d^2x\sqrt{|g|}\phi(x)^3}\Big)\rangle,\label{eq:dyson}
\ee
where $T_C$ denotes time ordering along the complex time (Keldysh) contour shown in Fig. \ref{fig:timecontour}.
\begin{figure}
\centering
\includegraphics[scale=1.3]{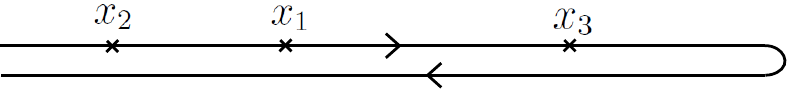}
\caption{\label{fig:timecontour} The complex time (Keldysh) contour for the in-in 3-point function.}
\end{figure}
Expanding (\ref{eq:dyson}) to first order in $\lambda$, and using Wick's theorem we obtain
\be
G_F^{(3)}(x_1,x_2,x_3)=-i\lambda\int_C d^2x \langle T_C(\phi(x)\phi(x_1))\rangle
\langle T_C(\phi(x)\phi(x_2))\rangle \langle T_C(\phi(x)\phi(x_3))\rangle.
\ee
Writing out the contour integral gives
\begin{align}
G_F^{(3)}(x_1,x_2,x_3)&=-i\lambda\int d^2x \langle T(\phi(x)\phi(x_1))\rangle
\langle T(\phi(x)\phi(x_2))\rangle \langle T(\phi(x)\phi(x_3))\rangle
\\
&+i\lambda\int d^2x \langle \phi(x)\phi(x_1)\rangle
\langle \phi(x)\phi(x_2)\rangle \langle \phi(x)\phi(x_3)\rangle,
\end{align}
where now the time integrals run from $\tau=-\infty$ to $\tau=+\infty$.

\begin{figure}
\centering
\includegraphics[scale=.9]{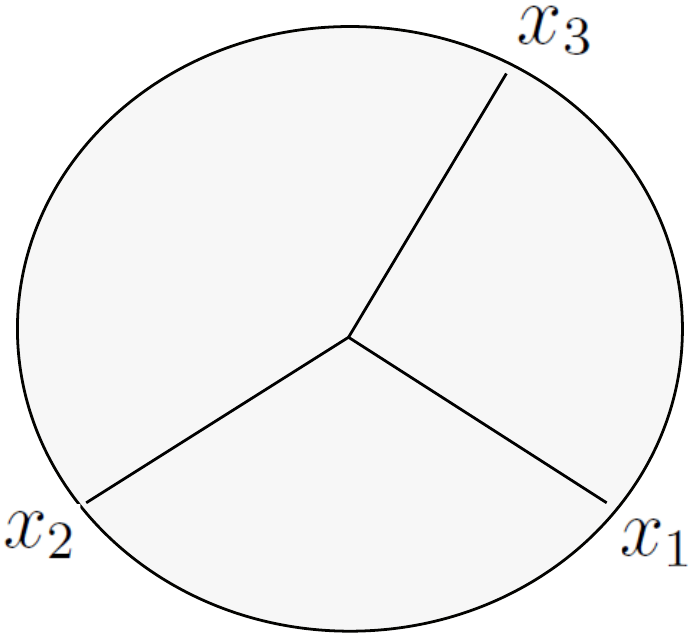}
\caption{\label{fig:3pt} The first order Feynman diagram contributing to the scalar three-point function.}
\end{figure}

Thus, the 3-point function is a sum of two Feynman diagrams of the form shown in Fig. \ref{fig:3pt}, with
the lines denoting in the first case time ordered two-point functions, and in the second case, Wightman two-point
functions. Using the known two-point functions gives
\begin{align}
G_F^{(3)}(x_1,x_2,x_3)&=-i\lambda \int d^2x\sqrt{|g|} \prod_{i=1}^{3}\frac{1}{\pi}\frac{z_i z}{-(\tau-\tau_i)^2+z^2+i\epsilon}\nonumber
\\
&+i\lambda \int d^2x\sqrt{|g|} \prod_{i=1}^{3}\frac{1}{\pi}\frac{z_i z}{-(\tau-\tau_i)^2+z^2+i(\tau-\tau_i)\epsilon}\label{eq:3ptintegral}
\end{align}
The boundary three-point function in the "extrapolate"  dictionary is given by
\be
G_{F,\partial}^{(3)}(t_1,t_2,t_3)=\lim_{z_i\rightarrow 0}(z_1 z_2 z_3)^{-1}G^{(3)}(x_1,x_2,x_3).
\ee
Thus, we finally obtain the bulk 3-point function as the sum of the following integrals
\begin{align}
G^{(3)}_{F,\partial}(\tau_1,\tau_2,\tau_3)&=-i\lambda \int dzd\tau\,z \prod_{i=1}^{3}\frac{1}{\pi}\frac{1}{-(\tau-\tau_i)^2+z^2+i\epsilon}\nonumber
\\
&+i\lambda \int dz dt\,z\sqrt{|g|} \prod_{i=1}^{3}\frac{1}{\pi}\frac{1}{-(\tau-\tau_i)^2+z^2+i(\tau-\tau_i)\epsilon}\label{eq:feynmanint}
\end{align}
We will first compute the $\tau$ integrals, from the residues at the poles of the
integrands. The Wightman two-point functions in the second line of (\ref{eq:feynmanint}) have poles at
\be
\tau=\tau_j\pm z+i\epsilon.
\ee
As the poles are all located in the upper part of the complex $\tau$ plane, we can close the integral contour from below without encountering any poles.
As the integrand vanishes as $\tau^{-6}$ at large $\tau$, the integral vanishes. Thus, the 3-point function reduces to the contribution from the time ordered
2-point functions, in the first line of (\ref{eq:feynmanint}). This is the expected result in the vacuum state, where the in-in and in-out formalisms are expected to agree. The time ordered two-point functions have poles at
\be
\tau=\tau_j\pm z\pm i\epsilon.
\ee
Performing the $\tau$ integral by closing the integral contour from the upper half complex $\tau$ plane gives
\be
G^{(3)}_{F,\partial}(\tau_1,\tau_2,\tau_3)=-\frac{\lambda}{4\pi^2}\int_0^{\infty}dz\sum_{j=1}^3\prod_{k\neq j}\frac{1}{t_{jk}}\frac{1}{z+\frac{1}{2}(\tau_{jk}+i\epsilon)},
\ee
where $\tau_{jk}=\tau_j-\tau_k$. The $z$ integrals are now elementary and can be performed using the identity
\be
\int_0^{\infty}\frac{dz}{(z+a)(z+b)}=\frac{1}{a-b}\log\Big(\frac{a}{b}\Big),
\ee
leading to
\be
G^{(3)}_{F,\partial}(\tau_1,\tau_2,\tau_3)=\frac{\lambda}{2\pi^2}\frac{1}{\tau_{12}\tau_{13}\tau_{23}}\log\Big(-1-i\epsilon \frac{\tau_1^2+\tau_2^2+\tau_3^2
-\tau_1 \tau_2-\tau_1 \tau_3-\tau_2 \tau_3}{\tau_{12}\tau_{13}\tau_{23}}\Big).
\ee
Using $\log(-1-i\delta)=-i\pi \textrm{sign}(\delta)$ for $\delta\rightarrow 0$, gives finally
\be
G^{(3)}_{F,\partial}(\tau_1,\tau_2,\tau_3)=-i\frac{\lambda}{2\pi}\frac{1}{|\tau_{12} \tau_{23}\tau_{13}|}.
\ee

\subsection{Vaidya 3-point functions}

Next, consider the 3-point function in the AdS-Vaidya spacetime. Using the real time Schwinger-Keldysh formalism, with the above two-point functions gives
the same integral expression as the ground state (\ref{eq:3ptintegral}), but now in the barred coordinates. One difference to the
vacuum case is that the integral over light-cone time now has an upper limit at $\bar{v}=2 z_H$ due to the coordinate relation (\ref{eq:coordinate2}).
It is easy to see that the integration region can be continued all the way to $\bar{v}=\infty$ without changing the value of the integral, as
in this region, the integrands in the first and second lines of (\ref{eq:3ptintegral}) identically cancel each other. This follows from unitarity,
as the Schwinger-Keldysh contour can be extended forwards in time without changing the result for the correlation function.
Thus, the same calculation of the integral in the vacuum 3-point function goes through and we obtain the near boundary 3-point function
\be
G^{(3)}_F=-i\frac{\lambda}{2\pi}\frac{\bar{z}_1\bar{z}_2\bar{z}_3}{|\bar{\tau}_{12}\bar{\tau}_{23}\bar{\tau}_{13}|}.
\ee
Now depending on whether $\bar{\tau}_j$ is before or after the collapse, we get different results. When all of the points are in the region $\bar{\tau}_j<0$,
we obtain the vacuum result. For $\bar{\tau}_1<0$, $\bar{\tau}_2<0$ and $\bar{\tau}_3>0$, we obtain using the extrapolate dictionary
\be
G^{(3)}_{F,\partial}=-i\frac{\lambda}{2\pi}\frac{1}{\cosh^2\frac{t_3}{l}}\frac{1}{|(t_1-t_2)(t_1-l\tanh\frac{t_3}{l})(t_2-l\tanh\frac{t_3}{l})|}.
\ee
On the other hand if two of the points $t_2$ and $t_3$ are located after the shell,  we obtain
\be
G^{(3)}_{F,\partial}=-i\frac{\lambda}{2\pi}\frac{1}{\cosh^2\frac{t_3}{l}\cosh^2\frac{t_2}{l}}\frac{1}{|(t_1-l\tanh\frac{t_2}{l})(t_1-l\tanh\frac{t_3}{l})(l\tanh\frac{t_2}{l}-l\tanh\frac{t_3}{l})|}.
\ee
And finally when all of the points are located after the shell, we obtain
\begin{align}
G^{(3)}_{F,\partial}&=-i\frac{\lambda}{2\pi}\frac{1}{\cosh^2\frac{t_3}{l}\cosh^2\frac{t_2}{l}\cosh^2\frac{t_1}{l}}\times\nonumber
\\
\times&\frac{1}{|(l\tanh\frac{t_1}{l}-l\tanh\frac{t_2}{l})(l\tanh\frac{t_1}{l}-l\tanh\frac{t_3}{l})(l\tanh\frac{t_2}{l}-l\tanh\frac{t_3}{l})|}.
\end{align}

\section{Four-point functions in finite temperature and with a quench}

In conformal quantum mechanics, it is almost as straightforward to derive the non-equilibrium four-point functions as the three-point functions. For completeness, and for possible
future reference, we end with this calculation. (For example, we expect that the structure of similar non-equilibrium
3- and 4-point functions in higher dimensional field theories reflects those of conformal quantum mechanics and can be reduced to them in an appropriate (equal space) limit. However,
we expect the computation to be more involved in higher dimensions.)
At zero temperature, for fields $\phi$ and $\tilde{\phi}$ with dimensions $\delta$ and $\tilde\delta$, respectively, the four-point function can be evaluated as (note, the corresponding expression in \cite{Jackiw:2012ur} contains a typo)
\be
\mbox{} \langle t_1|\phi(t_2)\tilde{\phi}(t_3) | t_4\rangle = \frac{\Gamma(2r_0)x^{r_0}\, _2F_1 (\delta,\tilde{\delta};2r_0;x)\langle 0|\phi(0)|0\rangle\langle 0|\tilde{\phi}(0)|0 \rangle}{2^{\delta+\tilde{\delta}+2r_0 }(t_{13})^{\tilde{\delta}-r_0}(t_{24})^{\delta-r_0}(t_{12})^{\delta+r_0}(t_{34})^{\tilde{\delta}+r_0}(t_{14})^{2r_0-\delta-\tilde{\delta}} } \ ,
\ee
where $t_{ij} = t_i-t_j$ and
\be
x = \frac{t_{12}t_{34}}{t_{13}t_{24}},
\ee
is the conformally invariant ratio familiar from the usual conformal field theories. In the above, we have set the scaling parameter $a$ to unity. Here, the $x$ dependent factors remind us of the model dependent functions of 4-point functions in conformal field theories. It is noteworthy that the above expression can be obtained by using only a single conformal block. Now, we do the change of variables to the thermal coordinates.
\be
x(\tau(t)) = \frac{\sinh(\frac{t_1-t_2}{l})\sinh(\frac{t_3-t_4}{l})}{\sinh(\frac{t_1-t_3}{l})\sinh(\frac{t_2-t_4}{l})} \equiv x_l
\ee
and
\bea
&&\langle t_1|\phi(t_2)\tilde{\phi}(t_3) | t_4\rangle = \nonumber \\
&& \frac{\langle 0|\phi(0)|0\rangle\langle 0|\tilde{\phi}(0)|0\rangle\Gamma(2r_0)x_l^{r_0}\, _2F_1 (\delta,\tilde{\delta};2r_0;x_l)2^{-\delta-\tilde{\delta}-2r_0 }l^{-2r_0-\delta-\tilde{\delta}}}{\sinh(\frac{t_{13}}{l})^{\tilde{\delta}-r_0}\sinh(\frac{t_{24}}{l})^{\delta-r_0}\sinh(\frac{t_{12}}{l})^{\delta+r_0}\sinh(\frac{t_{34}}{l})^{\tilde{\delta}+r_0}\sinh(\frac{t_{14}}{l})^{2r_0-\delta-\tilde{\delta}} } \ .
\eea

Now for the real thing, thermalization. Now, we set the quench moment at $t=0$. Also, $t_4>0$ and $t_1<0$. We consider three different cases. First, $t_2<0,t_3<0$:
\bea
&&\langle t_1|\phi(t_2)\tilde{\phi}(t_3) | t_4\rangle =  \frac{\langle 0|\phi(0)|0\rangle\langle 0|\tilde{\phi}(0)|0\rangle}{(t_{13})^{\tilde{\delta}-r_0}(t_{12})^{\delta+r_0}} \times \\
&&\mbox{} \frac{\Gamma(2r_0)x_q^{r_0}\, _2F_1 (\delta,\tilde{\delta};2r_0;x_q)2^{-\delta-\tilde{\delta}-2r_0 }}{(t_2\sinh(\frac{t_4}{l})-l\sinh(\frac{t_4}{l}))^{\delta-r_0}(t_3\cosh(\frac{t_4}{l})-l\sinh(\frac{t_4}{l}))^{\tilde{\delta}+r_0}(t_1\cosh(\frac{t_4}{l})-l\sinh(\frac{t_4}{l}))^{2r_0-\delta-\tilde{\delta}} }\ , \nonumber
\eea
with
\be
x_q \equiv \frac{t_{12}(\cosh(\frac{t_4}{l})t_3-l\sinh(\frac{t_4}{l}))}{t_{13}(\cosh(\frac{t_4}{l})t_2-l\sinh(\frac{t_4}{l}))} \ .
\ee

Then, $t_2<0,t_3>0$,
\bea
&&\langle t_1|\phi(t_2)\tilde{\phi}(t_3) | t_4\rangle = \frac{\langle 0|\phi(0)|0\rangle\langle 0|\tilde{\phi}(0)|0\rangle}{(t_{12})^{\delta+r_0}\sinh(\frac{t_{34}}{l})^{\tilde{\delta}+r_0}}\times \\
&&\frac{\Gamma(2r_0)x_q^{r_0}\, _2F_1 (\delta,\tilde{\delta};2r_0;x_q)2^{-\delta-\tilde{\delta}-2r_0 }l^{-\tilde{\delta}-r_0}}{(t_1\cosh(\frac{t_3}{l})-l\sinh(\frac{t_3}{l}))^{\tilde{\delta}-r_0}(t_2\sinh(\frac{t_4}{l})-l\sinh(\frac{t_4}{l}))^{\delta-r_0}   (t_1\cosh(\frac{t_4}{l})-l\sinh(\frac{t_4}{l}))^{2r_0-\delta-\tilde{\delta}} }\ , \nonumber
\eea
with
\be
x_q \equiv \frac{lt_{12}\sinh(\frac{t_{34}}{l})}{(\cosh(\frac{t_3}{l})t_1-l\sinh(\frac{t_3}{l}))(\cosh(\frac{t_4}{l})t_2-l\sinh(\frac{t_4}{l}))} \ .
\ee

Finally, when both $t_2,t_3>0$, we have
\bea
&&\langle t_1|\phi(t_2)\tilde{\phi}(t_3) | t_4\rangle =  \frac{\langle 0|\phi(0)|0\rangle\langle 0|\tilde{\phi}(0)|0\rangle}{(t_1\cosh(\frac{t_2}{l})-l\sinh(\frac{t_2}{l}))^{\delta+r_0}(t_1\cosh(\frac{t_3}{l})-l\sinh(\frac{t_3}{l}))^{\tilde{\delta}-r_0}}  \times  \\
&&\frac{\Gamma(2r_0)x_q^{r_0}\, _2F_1 (\delta,\tilde{\delta};2r_0;x_q)2^{-\delta-\tilde{\delta}-2r_0 }l^{-\tilde{\delta}-\delta}}{(t_1\cosh(\frac{t_4}{l})-l\sinh(\frac{t_4}{l}))^{2r_0-\delta-\tilde{\delta}}\sinh(\frac{t_{34}}{l})^{\tilde{\delta}+r_0}\sinh(\frac{t_{24}}{l})^{\delta-r_0}    }\ , \nonumber
\eea
where
\be
x_q \equiv \frac{(t_1\cosh(\frac{t_2}{l})-l\sinh(\frac{t_2}{l}))\sinh(\frac{t_{34}}{l})}{(t_1\cosh(\frac{t_3}{l})-l\sinh(\frac{t_3}{l}))\sinh(\frac{t_{24}}{l})} \ .
\ee
The results are of course what one expects, the structure of the four (and three)-point functions change in a systematic way at different points of time 
with respect to the quench moment. 

For further study, it would be interesting to include the backreaction corrections to the 3-point functions \cite{Almheiri:2014cka}, or to study how the 3- and 4-point functions
in CQM are recovered from higher dimensions. The non-equilibrium 2-point function agrees with the equal space limit of the 2-point function in 1+1 dimensions, in the limit
of large operator dimension when the 2-point function is well approximated by the geodesic approach. The analytical results
for the correlation functions would be interesting to compare with the equal space limits of $d$+1 correlation functions for $d>1$, because the blackening factor in the bulk black hole metric is
dimension dependent. It would also be interesting to see how the (semiclassical) limit of 4-point functions computed in higher dimensions \cite{Fitzpatrick:2014vua,Asplund:2014coa,Hijano:2015rla} reduce to the 4-point function
in CQM. 
\bigskip

\noindent
{\Large \bf Acknowledgments}

\medskip

JJ and EKV are in part supported by the Academy of Finland grant no
1268023. JJ is also in part supported by the U. Helsinki Graduate School PAPU. The  research of VK was supported by the European Research
Council  under  the  European  Union's  Seventh  Framework Programme (ERC Grant agreement 307955). EKV also thanks the Galileo Galilei Institute for its hospitality and
partial support, and  EKV and JJ 
thank the [Department of Energy's] Institute for Nuclear Theory at the University of Washington for its hospitality
and the Department of Energy for partial support, during the completion of this work.

\section*{Appendix}

\bigskip

We motivate the definition $O_{Vaidya}(t)$ in (\ref{OVaidya}) by checking that the time evolution of the state $|\tau\ket = O_P(\tau )|0\ket$ connects to that of $|t\ket
=O_{BH}(t)|0\ket$ after the quench. 

Recall that 
\be
 |\tau \ket = O_P(\tau )|0\ket = e^{i\tau H_P}e^{aH_P}|0\ket
\ee
where the second exponential prepares the state $|\tau=0\ket$ and the first factor $e^{i\tau H_P}$ continues the time evolution (also to earlier times). The state $|\tau=0\ket
=|t=0\ket$ at the quench. After the quench, the time evolution continues with the new time translation generator $H_{BH}$, so it should evaluate to the state
\be
 |t \ket' \equiv O_{BH}(t )|0\ket = e^{it H_{BH}}e^{aH_P}|0\ket \ .
\ee
On the other hand, we had defined the state $|t\ket$ to be
\be
|t\ket = \left(\frac{d\tau}{dt} \right)^{r_0} |\tau = \tau (t)\ket = \left[ \cosh (t/l)\right]^{-2r_0} |\tau = l \tanh (t/l)\ket 
\ee
where $l=2u_H$.
Thus, we need to show that the two states are the same, $|t\ket'=|t\ket$. To establish that, we study how they time evolve. On one hand,
\begin{eqnarray}
 -i\partial_t |t\ket &=& \left[ i\frac{2r_0}{l} \sinh (t/l) [\cosh (t/l)]^{-1} + \cosh^{-2}(t/l) H_P \right] |t\ket \nonumber \\
\mbox{} &=& \cosh^{-2}(t/l) \left[ H_P + i\frac{2r_0}{l}\cosh (t/l)\sinh (t/l)\right] |t\ket .
\end{eqnarray}
On the other hand,
\be
 -i\partial_t |t\ket' = H_{BH}|t\ket' = (H_P+\gamma K) |t\ket' \ ,
\ee
where $\gamma = -1/l^2$.
We then show that the above two first order equations are in fact identical. Since the initial state $|t=\tau=0\ket=e^{aH_p}|0\ket$  is the same, that ensures that 
$|t\ket' = |t\ket$.  

Starting from 
\begin{eqnarray}
  K|t\ket &=& -i\left( \tau^2 \partial_\tau + 2r_0 \tau \right) |t \ket = \left(\tau^2 H_P - i2r_0\tau \right)  |t \ket
  \nonumber \\
\mbox{} &=& \left[ \frac{1}{\gamma} \left( \cosh^{-2} (t/l) -1\right)H_P + \frac{i2r_0}{\gamma l} \tanh (t/l) \right] |t\ket 
\end{eqnarray}
we obtain
\begin{eqnarray}
(H_P+\gamma K) |t\ket = \left[\cosh^{-2}(t/l) H_P + \frac{i2r_0}{l}\tanh (t/l) \right] |t\ket = -i\partial_t |t\ket
\end{eqnarray}
so that $|t\ket$ indeed satisfies the same first order equation as $|t\ket'$.

\bigskip


\begin{thebibliography}{10}


\bibitem{daff}
  V.~de Alfaro, S.~Fubini and G.~Furlan,
  ``Conformal Invariance in Quantum Mechanics,''
  Nuovo Cim.\ A {\bf 34}, 569 (1976).
\bibitem{Maldacena:1998uz} 
  J.~M.~Maldacena, J.~Michelson and A.~Strominger,
  ``Anti-de Sitter fragmentation,''
  JHEP {\bf 9902}, 011 (1999)
  [hep-th/9812073].
\bibitem{Almheiri:2014cka} 
  A.~Almheiri and J.~Polchinski,
  ``Models of AdS$_2$  Backreaction and Holography,''
  arXiv:1402.6334 [hep-th].
\bibitem{Banks:1998dd} 
  T.~Banks, M.~R.~Douglas, G.~T.~Horowitz and E.~J.~Martinec,
  ``AdS dynamics from conformal field theory,''
  hep-th/9808016.
\bibitem{Hubeny:2007xt} 
  V.~E.~Hubeny, M.~Rangamani and T.~Takayanagi,
  ``A Covariant holographic entanglement entropy proposal,''
  JHEP {\bf 0707}, 062 (2007)
  [arXiv:0705.0016 [hep-th]].
\bibitem{AbajoArrastia:2010yt} 
  J.~Abajo-Arrastia, J.~Aparicio and E.~Lopez,
  ``Holographic Evolution of Entanglement Entropy,''
  JHEP {\bf 1011}, 149 (2010)
  [arXiv:1006.4090 [hep-th]].
\bibitem{Albash:2010mv} 
  T.~Albash and C.~V.~Johnson,
  ``Evolution of Holographic Entanglement Entropy after Thermal and Electromagnetic Quenches,''
  New J.\ Phys.\  {\bf 13}, 045017 (2011)
  [arXiv:1008.3027 [hep-th]].
\bibitem{Balasubramanian:2010ce} 
  V.~Balasubramanian {\it et al.},
  ``Thermalization of Strongly Coupled Field Theories,''
  Phys.\ Rev.\ Lett.\  {\bf 106}, 191601 (2011)
  [arXiv:1012.4753 [hep-th]].
\bibitem{Balasubramanian:2011ur} 
  V.~Balasubramanian {\it et al.},
  ``Holographic Thermalization,''
  Phys.\ Rev.\ D {\bf 84}, 026010 (2011)
  [arXiv:1103.2683 [hep-th]].
\bibitem{cjps} 
  C.~Chamon, R.~Jackiw, S.~Y.~Pi and L.~Santos,
  ``Conformal quantum mechanics as the CFT$_1$ dual to AdS$_2$,''
  Phys.\ Lett.\ B {\bf 701}, 503 (2011)
  [arXiv:1106.0726 [hep-th]].
\bibitem{3pt4pt}
P.~Arnold and D.~Vaman,
  JHEP {\bf 1010}, 099 (2010)
  doi:10.1007/JHEP10(2010)099
  [arXiv:1008.4023 [hep-th]]; 
 P.~Arnold and D.~Vaman,
  JHEP {\bf 1104} (2011) 027
  doi:10.1007/JHEP04(2011)027
  [arXiv:1101.2689 [hep-th]];
 P.~Arnold and D.~Vaman,
  J.\ Phys.\ G {\bf 38}, 124175 (2011)
  doi:10.1088/0954-3899/38/12/124175
  [arXiv:1106.1680 [hep-th]];
 P.~Arnold, D.~Vaman, C.~Wu and W.~Xiao,
  JHEP {\bf 1110}, 033 (2011)
  doi:10.1007/JHEP10(2011)033
  [arXiv:1105.4645 [hep-th]];
 O.~Saremi and K.~A.~Sohrabi,
  JHEP {\bf 1111}, 147 (2011)
  doi:10.1007/JHEP11(2011)147
  [arXiv:1105.4870 [hep-th]].
\bibitem{Fitzpatrick:2014vua} 
  A.~L.~Fitzpatrick, J.~Kaplan and M.~T.~Walters,
  ``Universality of Long-Distance AdS Physics from the CFT Bootstrap,''
  JHEP {\bf 1408}, 145 (2014)
  [arXiv:1403.6829 [hep-th]].
\bibitem{Asplund:2014coa} 
  C.~T.~Asplund, A.~Bernamonti, F.~Galli and T.~Hartman,
  ``Holographic Entanglement Entropy from 2d CFT: Heavy States and Local Quenches,''
  JHEP {\bf 1502}, 171 (2015)
  [arXiv:1410.1392 [hep-th]].
\bibitem{Hijano:2015rla} 
  E.~Hijano, P.~Kraus and R.~Snively,
  ``Worldline approach to semi-classical conformal blocks,''
  JHEP {\bf 1507}, 131 (2015)
  [arXiv:1501.02260 [hep-th]].
\bibitem{Jackiw:2012ur}
  R.~Jackiw and S.-Y.~Pi,
  ``Conformal Blocks for the 4-Point Function in Conformal Quantum Mechanics,''
  Phys.\ Rev.\ D {\bf 86}, 045017 (2012)
  [Phys.\ Rev.\ D {\bf 86}, 089905 (2012)]
  [arXiv:1205.0443 [hep-th]].
\bibitem{hartnoll} 
  D.~Anninos, S.~A.~Hartnoll and D.~M.~Hofman,
  ``Static Patch Solipsism: Conformal Symmetry of the de Sitter Worldline,''
  Class.\ Quant.\ Grav.\  {\bf 29}, 075002 (2012)
  [arXiv:1109.4942 [hep-th]].
\bibitem{nakayama} 
  R.~Nakayama,
  ``The World-Line Quantum Mechanics Model at Finite Temperature which is Dual to the Static Patch Observer in de Sitter Space,''
  Prog.\ Theor.\ Phys.\  {\bf 127}, 393 (2012)
  [arXiv:1112.1267 [hep-th]].

\bibitem{spradlin-strominger}
 M.~Spradlin and A.~Stromiger,
 ``Vacuum States for AdS$_2$ Black Holes,''
 JHEP {\bf 9911}, 021 (1999)
 [hep-th/9904143].

\bibitem{Ho:2004qp} 
  P.~M.~Ho,
  JHEP {\bf 0405}, 008 (2004)
  doi:10.1088/1126-6708/2004/05/008
  [hep-th/0401167].

\bibitem{Ebrahim:2010ra}
  H.~Ebrahim and M.~Headrick,
  ``Instantaneous Thermalization in Holographic Plasmas,''
  arXiv:1010.5443 [hep-th].
\bibitem{Keranen:2014lna} 
  V.~Keranen and P.~Kleinert,
  ``Non-equilibrium scalar two point functions in AdS/CFT,''
  JHEP {\bf 1504}, 119 (2015)
  [arXiv:1412.2806 [hep-th]].
\bibitem{spectral}
  V.~Balasubramanian, A.~Bernamonti, B.~Craps, V.~Ker\"anen, E.~Keski-Vakkuri, B.~M\"uller, L.~Thorlacius and J.~Vanhoof,
  ``Thermalization of the spectral function in strongly coupled two dimensional conformal field theories,''
  JHEP {\bf 1304}, 069 (2013)
  [arXiv:1212.6066 [hep-th]].
\bibitem{Calabrese:2009qy} 
  P.~Calabrese and J.~Cardy,
  ``Entanglement entropy and conformal field theory,''
  J.\ Phys.\ A {\bf 42}, 504005 (2009)
  [arXiv:0905.4013 [cond-mat.stat-mech]].
\end{thebibliography}

\providecommand{\href}[2]{#2}\begingroup\raggedright

\end{document}